\documentclass[fleqn,10pt]{SelfArx} % Document font size and equations flushed left

\definecolor{color1}{RGB}{0,0,90} % Color of the article title and sections
\definecolor{color2}{RGB}{0,20,20} % Color of the boxes behind the abstract and headings

\usepackage{hyperref} % Required for hyperlinks
\hypersetup{hidelinks,colorlinks,breaklinks=true,urlcolor=color2,citecolor=color1,linkcolor=color1,bookmarksopen=false,pdftitle={Title},pdfauthor={Author},urlcolor=blue}

\usepackage{graphicx}
\usepackage[square]{natbib}
\usepackage{amsmath}
\usepackage{multirow}
\usepackage{subfigure}

\newcommand{\n}[1]{\mathrm{#1}}

\JournalInfo{Published in Journal of Applied Physics, Vol. 109 (1), 013915, 2011} % Journal information
\Archive{\href{http://dx.doi.org/10.1063/1.3525646}{DOI: 10.1063/1.3525646}} % Additional notes (e.g. copyright, DOI, review/research article)

\PaperTitle{The ideal dimensions of a Halbach cylinder of finite length} % Article title

\Authors{R. Bj\o{}rk} % Authors
\affiliation{\textit{Department of Energy Conversion and Storage, Technical University of Denmark - DTU, Frederiksborgvej 399, DK-4000 Roskilde, Denmark}} % Author affiliation
\affiliation{*\textbf{Corresponding author}: rabj@dtu.dk} % Corresponding author

\Keywords{} % Keywords - if you don't want any simply remove all the text between the curly brackets
\newcommand{\keywordname}{Keywords} % Defines the keywords heading name

\Abstract{In this paper the smallest or optimal dimensions of a Halbach cylinder of a finite length for a given sample volume and desired flux density are determined using numerical modeling and parameter variation. A sample volume that is centered in and shaped as the Halbach cylinder bore but with a possible shorter length is considered. The external radius and the length of the Halbach cylinder with the smallest possible dimensions are found as a function of a desired internal radius, length of the sample volume and mean flux density. It is shown that the optimal ratio between the outer and inner radius of the Halbach cylinder does not depend on the length of the sample volume. Finally, the efficiency of a finite length Halbach cylinder is considered and compared with the case of a cylinder of infinite length. The most efficient dimensions for a Halbach cylinder are found and it is shown that the efficiency increases slowly with the length of the cylinder.}

\begin{document}

\flushbottom % Makes all text pages the same height

\maketitle % Print the title and abstract box

%\tableofcontents % Print the contents section

\thispagestyle{empty} % Removes page numbering from the first page

\section{Introduction}\label{Sec.Introduction}
Permanent magnet assemblies that produce a strong homogenous magnetic field in a sample volume are becoming of general interest to the scientific community as these have an increasing number of applications. Of special interest is the Halbach cylinder (also known as a hole cylinder permanent magnet array (HCPMA)) which is a hollow permanent magnet cylinder with a remanent flux density at any point that varies continuously as
\begin{eqnarray}
B_{\mathrm{rem},r}    &=& B_{\mathrm{rem}}\; \textrm{cos}(p\phi) \nonumber\\
B_{\mathrm{rem},\phi} &=& B_{\mathrm{rem}}\; \textrm{sin}(p\phi)\;,\label{Eq.Halbach_magnetization}
\end{eqnarray}
where $B_{\mathrm{rem}}$ is the magnitude of the remanent flux density and $p$ is an integer \citep{Mallinson_1973,Halbach_1980}. Subscript $r$ denotes the radial component of the remanence and subscript $\phi$ the tangential component. A positive value of $p$ produces a field that is directed into the cylinder bore and a negative value produces a field that is directed outwards from the cylinder.

Often, as is also the case here, the Halbach cylinder with $p=1$ is considered and will here simply be referred to as the Halbach cylinder. This design generates a completely homogenous magnetic field in the cylinder bore with a flux density that for an infinitely long cylinder is given by
\begin{eqnarray} \label{Eq.Halbach_flux_density}
B = B_\n{rem}\n{ln}\left(\frac{r_\n{o}}{r_\n{i}}\right)~,
\end{eqnarray}
where $r_\n{o}$ and $r_\n{i}$ are the outer and inner radii of the cylinder, respectively. Thus the Halbach cylinder is able to generate a homogeneous flux density that is larger than the remanence of the permanent magnet used in the design. Also, the Halbach cylinder can, because it is well defined geometrically, easily be adapted to an experimental setup.

The Halbach cylinder has previously been used in a number of applications such as magnetic refrigeration devices \cite{Tura_2007,Bjoerk_2010b}, nuclear magnetic resonance (NMR) apparatus \citep{Moresi_2003,Appelt_2006} and accelerator magnets \citep{Sullivan_1998,Lim_2005}.

The magnetic field distribution for a Halbach cylinder of infinite length have previously been investigated in detail  \citep{Zhu_1993,Atallah_1997,Peng_2003,Xia_2004,Bjoerk_2010a}, but Halbach cylinders of a finite length have not been considered in detail. The reduction in flux density due to a finite length Halbach cylinder have been considered mostly for cylinders with a single fixed length \cite{Mhiochain_1999,Xu_2004}. The ideal outer radius and length for a Halbach cylinder with a fixed inner radius has also been presented \cite{Bjoerk_2008}, but here the flux density in the complete cylinder bore and not a specific sample volume was considered. Finally, an analytical formula for the magnetic flux density of a Halbach cylinder of any given length has been derived, however this formula is extremely complicated, making it impractical for direct application \cite{Mhiochain_1999}.

\begin{figure}[!t]
  \centering
   \includegraphics[width=1\columnwidth]{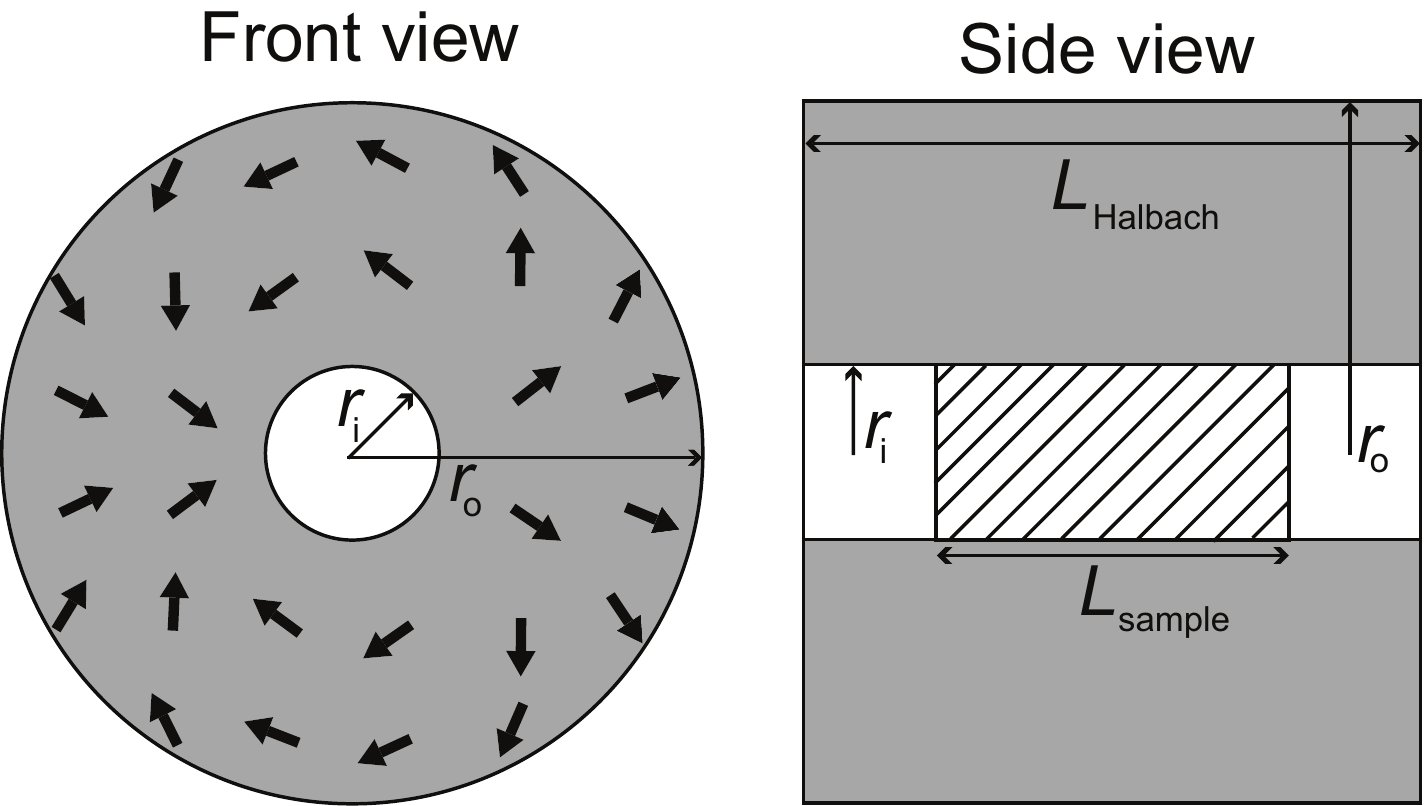}
      \caption{An illustration of the Halbach cylinder. On the front view the direction of the magnetization is shown as arrows. On the side view the cylinder has been made transparent so that the sample volume can be seen. The sample volume has been hashed.}
      \label{Fig.Halbach_illustration_fixed_bore}
\end{figure}

Here, a Halbach cylinder of a finite length, $L_\n{Halbach}$, is considered and the optimal dimensions for this design are investigated using numerical modeling and parameter variation of the dimensions of the Halbach cylinder. The optimal dimensions are those where the volume of the magnet is the smallest possible for a desired flux density and sample volume. A desired sample volume, centered in and shaped as the Halbach cylinder bore but with a possible shorter length, $L_\n{sample}$, is considered. The aim is to determine the external radius and the length of the Halbach cylinder with the smallest possible dimensions given a desired internal radius, length of the sample volume and mean flux density in this volume. The most efficient Halbach cylinder of a finite length, defined using a figure of merit, is also considered. An illustration of the Halbach cylinder and the sample volume is shown in Fig. \ref{Fig.Halbach_illustration_fixed_bore}.

For most applications a flux density of $0.5-2.5$ T is needed and thus this range will be considered here. Creating a flux density higher than 2.5 T requires that one carefully consider the intrinsic coercivity of the permanent magnets used, as the reverse component of the magnetic field can exceed the intrinsic coercivity in parts of the Halbach cylinder near the cylinder bore which will lead to a reversal of the direction of magnetization of parts of the magnet \cite{Bjoerk_2008,Bloch_1998}.

A Halbach cylinder where the direction of the remanence is given by Eq. (\ref{Eq.Halbach_magnetization}) is considered. In real world assemblies the Halbach cylinder is often segmented into blocks each with their own constant direction of magnetization. The effect of this segmentation has been considered elsewhere \cite{Halbach_1980,Mhiochain_1999,Bjoerk_2008} and will not be considered here. For a sixteen segmented Halbach cylinder of infinite length the flux density is reduced to 97.5\% of the value of a Halbach cylinder with a continuously rotating remanence.

\begin{figure}[!t]
  \centering
   \includegraphics[width=1\columnwidth]{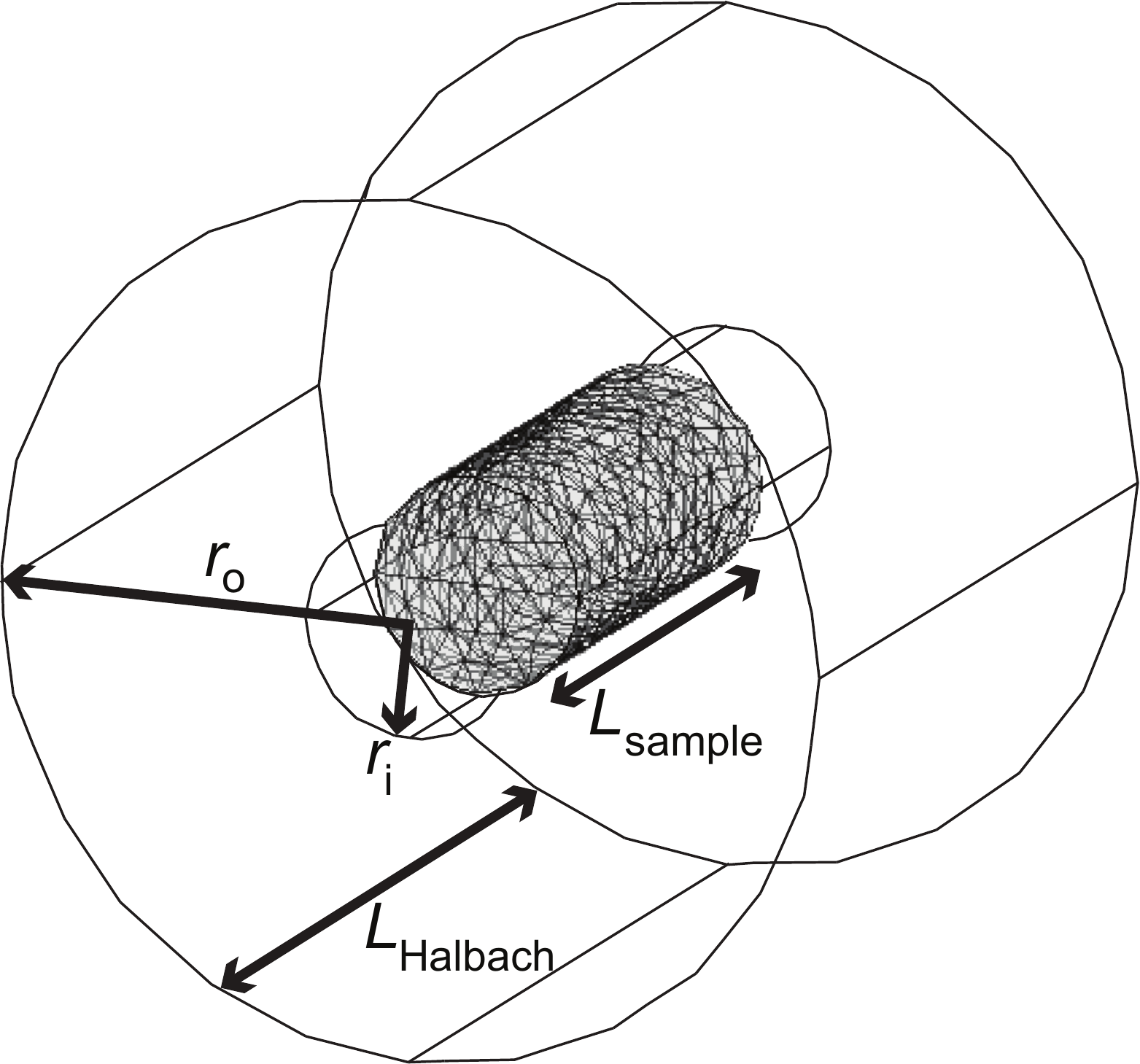}
      \caption{A 3D illustration of a transparent Halbach cylinder where the internal radius, external radius and the length of the Halbach cylinder and the sample volume has been indicated. For this illustration these have the values $r_\n{i} = 10$ mm, $r_\n{o} = 40$ mm, $L_\n{Halbach} = 70$ mm and $L_\n{sample} = 50$ mm. The particular finite element mesh used in the sample volume is also shown.}
      \label{Fig.Halbach_mesh_3D}
\end{figure}

In order to determine the flux density of a given Halbach cylinder the system is modeled numerically. This is done using the commercially available finite element multiphysics program \emph{Comsol Multiphysics}, version 3.5a. The Comsol Multiphysics code has previously been validated through a number of NAFEMS (National Agency for Finite Element Methods and Standards) benchmark studies \citep{Comsol_2005}. The resolution of the mesh used for the simulations presented in this paper are chosen such that the result do not depend on the mesh size. To ensure this the mesh is refined in the sample volume to increase the number of mesh elements there. An illustration of the mesh used is shown in Fig. \ref{Fig.Halbach_mesh_3D}. As the size of the computational volume as well as the size of the Halbach cylinder changes for the different dimensions of the sample volume and Halbach cylinder the number of mesh elements changes as well.

The system modeled is a static problem in magnetism, i.e. magnetostatics. The equation solved is the magnetic scalar potential equation, which is a Poisson equation,
\begin{eqnarray}
-\nabla{}\cdot{}(\mu{}_{0}\mu{}_{r}\nabla{}V_\mathrm{m}-\mathbf{B}_{\mathrm{rem}})=0~,\label{Eq.Numerical_Magnetism}
\end{eqnarray}
where $\mu{}_{0}$ is the permeability of free space, $\mu{}_{r}$ is the relative permeability, which in all cases is assumed to be isotropic, and $V_\mathrm{m}$ is the magnetic scalar potential from which the magnetic field can be found as $-\nabla{}V_\mathrm{m} = \mathbf{H}$. The remanent flux density, $\mathbf{B}_{\mathrm{rem}}$, is known as a function of position and can be used as an initial condition to solve the system.

The above equation is solved on a finite element mesh and the solver used is \emph{Pardiso} which is a parallel sparse direct linear solver \citep{Schenk_2001,Schenk_2002}. Boundary conditions are chosen such that the boundaries of the computational volume, which is many times larger than the simulated magnetic structure, are magnetically insulating, i.e. $\hat{\mathbf{n}}\cdot{}\mathbf{B}=0$, where $\hat{\mathbf{n}}$ is the surface normal, while for all other (internal) boundaries the magnetic boundary conditions apply, i.e. the perpendicular component of $\mathbf{B}$ and the parallel components of $\mathbf{H}$ are continuous across boundaries. The computational volume is always chosen large enough that the insulating boundaries do not affect the calculations.

The Halbach cylinder is modeled with a relative constant isotropic permeability of $\mu_r = 1$. The direction of the remanence in each point is fixed, which is only valid if the anisotropy field is very large and if the component of $\mathbf{H}$ that is parallel and opposite to $\mathbf{B}_{\mathrm{rem}}$ is less than the intrinsic coercivity. For an NdFeB magnet, which is typically used for construction of a Halbach cylinder, the relative permeability is 1.05, the intrinsic coercivity can be as high as 3.2 T and the anisotropy field has a value of 8 T \citep{Zimmermann_1993}, thus this modeling approach is justified.

\section{Optimal dimensions of a Halbach cylinder}
As previously mentioned the aim is to determine the optimal dimensions of a Halbach cylinder, i.e. the external radius and the length of the smallest Halbach cylinder, given a desired internal radius, length of the sample volume and mean flux density in this volume. This is done through a numerical modeling of the Halbach cylinder design where the dimensions of the cylinder has been varied. For this variation the inner radius was fixed at 10 mm. The outer radius was varied from 11 mm to 120 mm in steps of 1 mm, the length of the sample volume was varied from 10 mm to 80 mm in steps of 1 mm and the length of the Halbach cylinder was varied from the length of the sample volume to 120 mm in steps of 1 mm. This corresponds to a change in the ratio between the outer and inner radius of 1.1 to 12 and a ratio between the length of the Halbach cylinder and the inner radius from a minimum value of 1 to 12. Note that the sample volume can be smaller than the Halbach cylinder bore. The total number of parameter variations is 593561. Such a large parameter space requires a substantial calculation time, especially considering that a single simulation typically takes of the order of 30 seconds to compute. Therefore the simulations have been calculated on a number of powerful computers, each running multiple simulations at once. For all configurations the mean flux density in the sample volume, denoted $\langle B \rangle$, was calculated. The flux density in the center of the bore can be calculated analytically \cite{Zijlstra_1985}, but no analytical expression exist for the mean value of the flux density in the sample volume.

It is very important to note that a magnetostatic problem is scale invariant, i.e. if all dimensions are scaled by the same factor the magnetic flux density in a given point will be the same if the coordinates of this point is scaled as well. This means that quantities such as the average value of the magnetic flux density in the sample volume will be the same, as long as the magnet design is scaled appropriately. Therefore all results here are presented in dimensionless quantities that can be scaled to obtain the dimensions for a desired inner radius and length of the sample volume.

The remanence was fixed at 1.4 T. However, it is remarked that because the modeled Halbach cylinder permanent magnet has a relative permeability of one, the magnetostatic problem of calculating the flux density is linear in the remanence. This means that all magnetic flux densities reported here can simply be scaled by the desired remanence, and therefore all results are reported as function of the flux density divided by the remanence.

\section{Results}
\begin{figure}[!t]
  \centering
   \includegraphics[width=1\columnwidth]{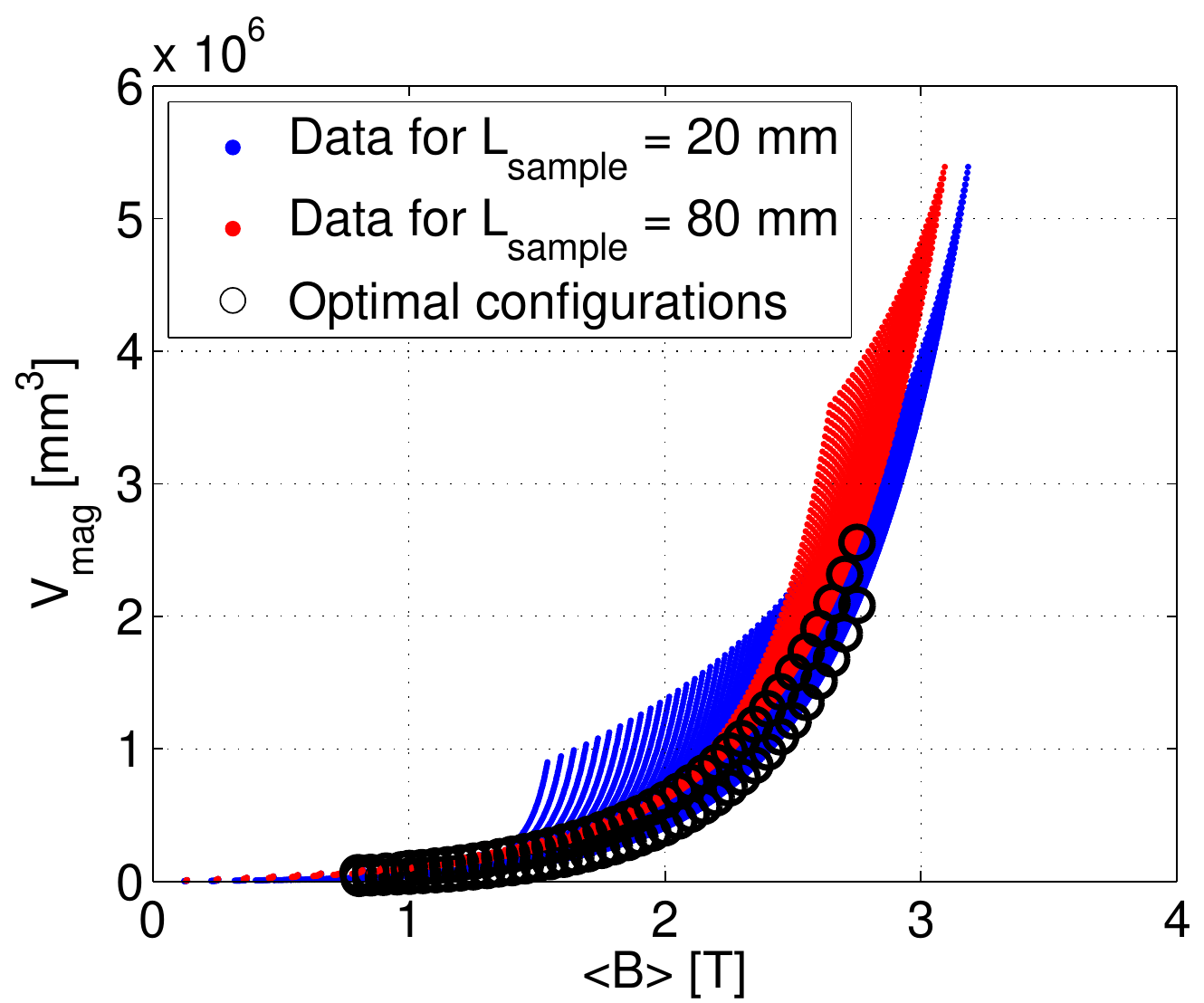}
      \caption{The volume of the magnet as a function of the mean flux density for a length of the sample volume of $L_\n{sample} = 20$ mm and 80 mm. The optimal configurations, i.e. those with the lowest value of $V_\n{mag}$ for a given value of $\langle B \rangle$, for both cases have been indicated.}
      \label{Fig.V_mag_B_optimal_points}
\end{figure}

For each value of $L_\n{sample}$ the optimal configurations, i.e. the configurations with the smallest volume of the magnet at a given mean flux density, has been found, similarly to the procedure described in Ref. \cite{Bjoerk_2008}. The optimal configurations have been found from 0.8 T to 2.8 T in steps of 0.05 T, and for the two specific cases of $L_\n{sample} = 20$ and 80 mm these are shown in Fig. \ref{Fig.V_mag_B_optimal_points}.

\begin{figure}[!t]
  \centering
   \includegraphics[width=1\columnwidth]{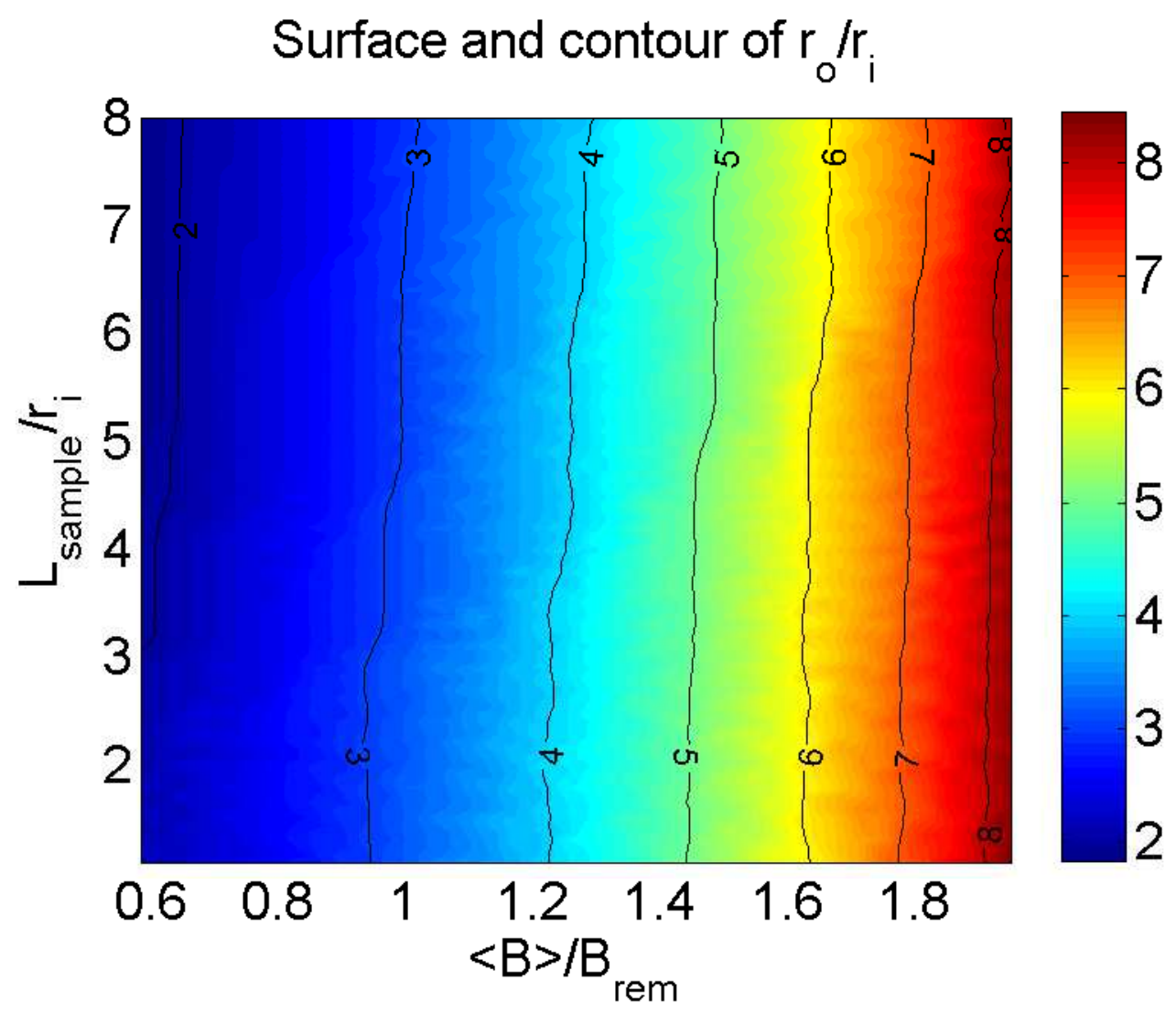}
      \caption{A contour and surface map of the ratio between the outer and inner radius as a function of the ratio between the mean flux density and the remanence and the ratio between the length of the sample volume and the inner radius for the optimal configurations, i.e. those that generate the desired flux density using the least amount of magnet material.}
      \label{Fig.r_o_over_r_i_surface_image}
\end{figure}

\begin{figure}[!t]
  \centering
   \includegraphics[width=1\columnwidth]{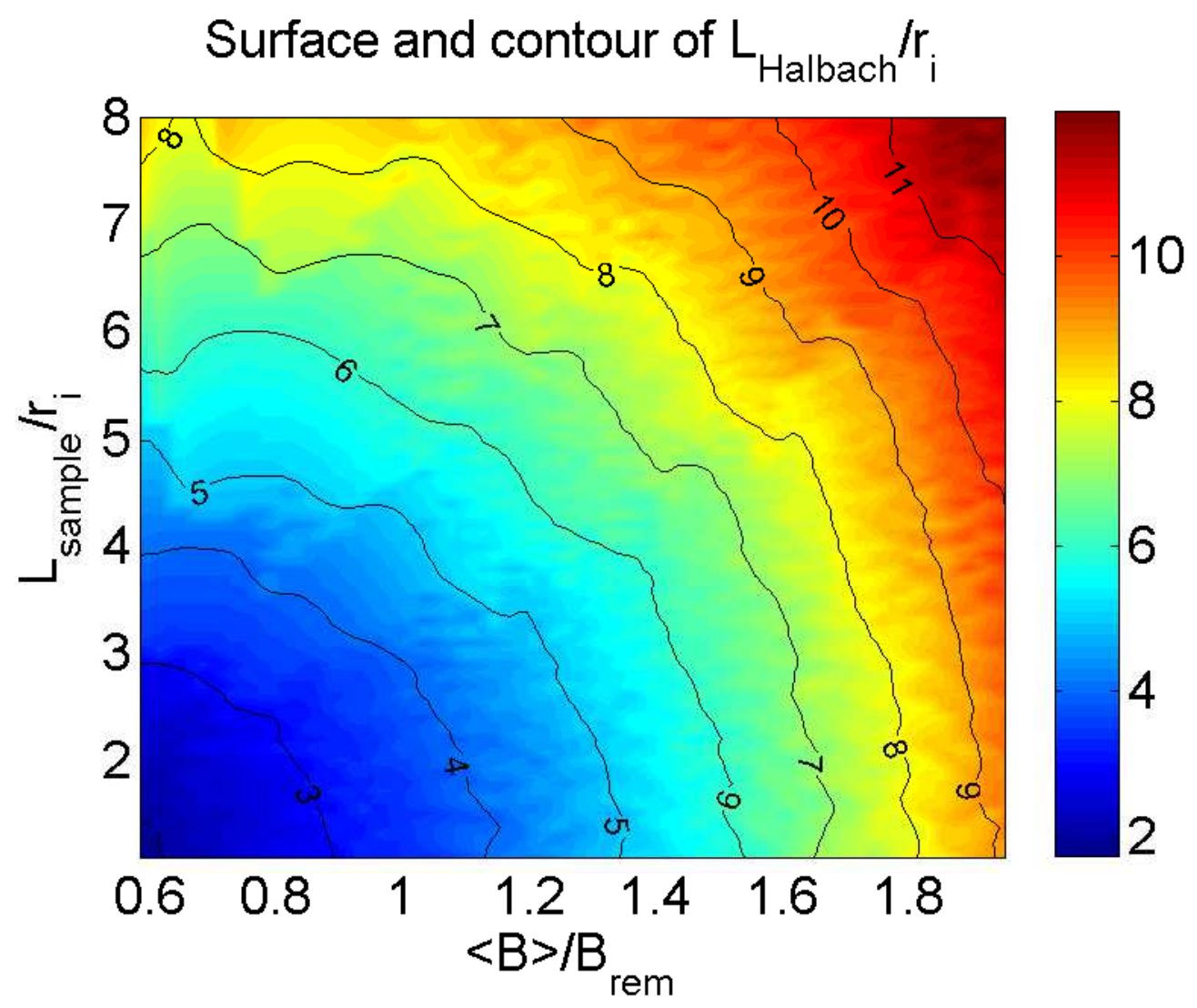}
      \caption{A contour and surface map of the ratio between the length of the Halbach cylinder and the inner radius as a function of the ratio between the mean flux density and the remanence and the ratio between the length of the sample volume and the inner radius for the optimal configurations, i.e. those that generate the desired flux density using the least amount of magnet material.}
      \label{Fig.L_halbach_over_r_i_surface_image}
\end{figure}

The dimensions of these optimal configurations i.e. the outer radius and the length of the Halbach cylinder can thus be investigated as function of $L_\n{sample}$ and $\langle B \rangle$. In Figs. \ref{Fig.r_o_over_r_i_surface_image} and \ref{Fig.L_halbach_over_r_i_surface_image} the outer radius and the length, respectively, both in units of the inner radius, is shown for the optimal configurations as a function of the flux density and the length of the sample volume, each normalized with the remanence and the inner radius, respectively. Thus the optimal dimensions of a Halbach cylinder with a desired inner radius, length of the sample volume and mean flux density can be found directly from these figures.

As an example consider a desired sample volume with a radius of 20 mm, a length of 80 mm and a mean flux density of 1 T. For the construction of the Halbach cylinder magnets with a remanence of 1.2 T are considered. Based on Fig. \ref{Fig.r_o_over_r_i_surface_image} the outer radius of the cylinder must be $\sim 54$ mm and based on Fig. \ref{Fig.L_halbach_over_r_i_surface_image} the length of the Halbach cylinder must be $\sim 90$ mm.

From Fig. \ref{Fig.r_o_over_r_i_surface_image} it is seen that the ratio between the outer and inner radius does not depend on the length of the sample volume. This allows for easy dimensioning of the optimal Halbach cylinder as the ratio $r_\n{o}/r_\n{i}$ can thus be chosen directly from the desired flux density.

In Fig. \ref{Fig.L_halbach_over_r_i_surface_image} the length of the Halbach cylinder is seen to depend on both the desired flux density and the length of the sample volume. For small values of the flux density the length of the Halbach cylinder is close to the length of the sample volume, while for larger flux densities it is advantageous to make the Halbach cylinder longer then the desired length of the sample volume. The unevenness of the contour lines in Fig. \ref{Fig.L_halbach_over_r_i_surface_image} are due to the finite steps of the parameter variation as well as the way the optimal configurations are determined, i.e. by searching a finite data set for an absolute minimum value.

\begin{figure}[!t]
  \centering
   \includegraphics[width=1\columnwidth]{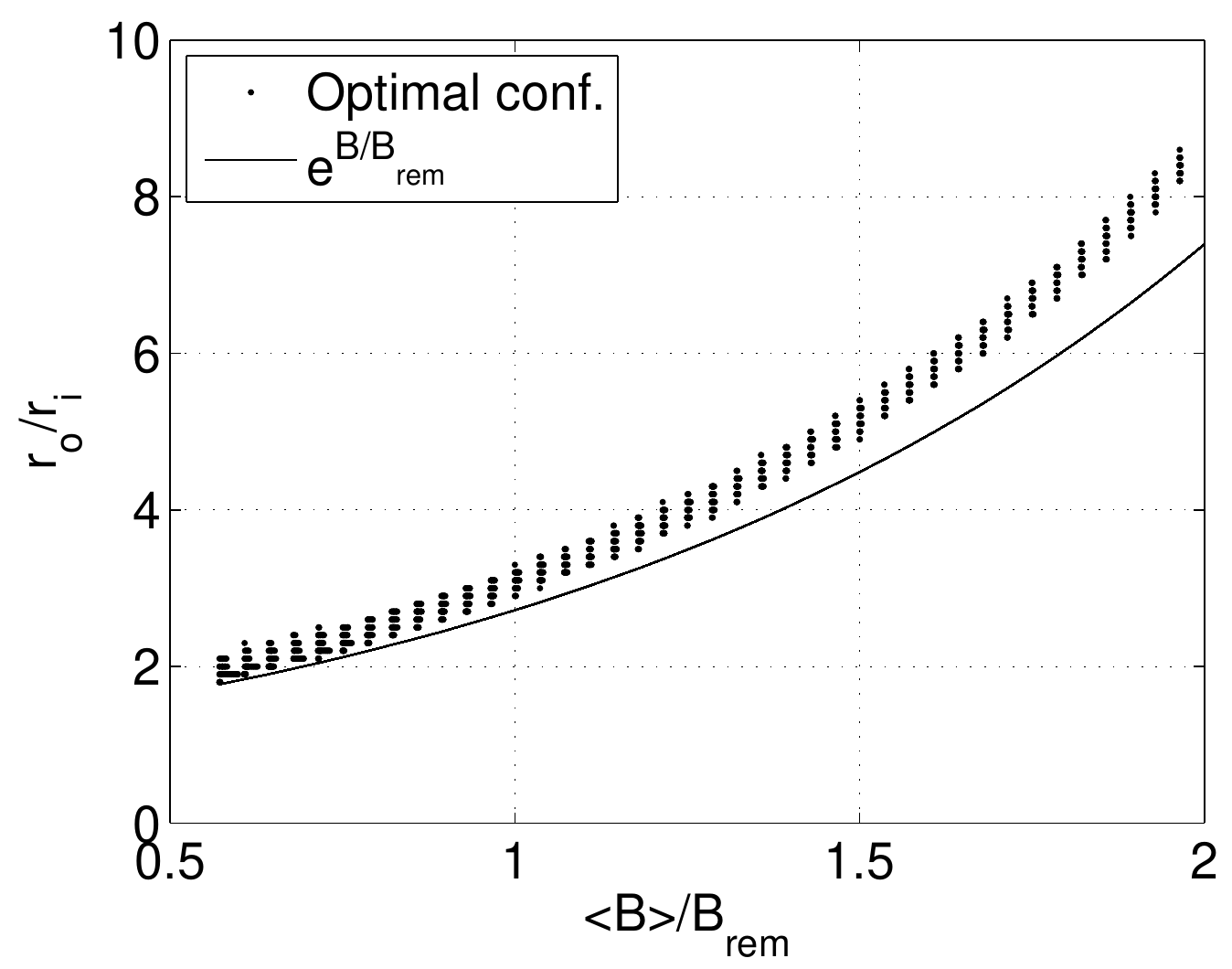}
      \caption{The ratio between the outer and the inner radius as a function of the ratio between the flux density and the remanence for the optimal configurations, i.e. the configuration with the smallest volume of the magnets for a given mean flux density for all values of $L_\mathrm{sample}/r_\mathrm{i}$. The ratio for a infinite cylinder, given by $e^{B/B_\n{rem}}$, is also shown.}
      \label{Fig.r_o_function_of_B}
\end{figure}

As discussed the ratio between the outer and inner radius does not depend on the length of the sample volume. This can also be seen in Fig. \ref{Fig.r_o_function_of_B}, which is a collapse of Fig. \ref{Fig.r_o_over_r_i_surface_image} along the $L_\n{sample}/r_\n{i}$ axis. For a Halbach cylinder of infinite length the ratio between the outer and the inner radius can be calculated as a function of the ratio between the flux density and the remanence as
\begin{eqnarray}
\frac{r_\mathrm{o}}{r_\mathrm{i}} = e^{B/B_\mathrm{rem}}~.
\end{eqnarray}
This is also shown in Fig. \ref{Fig.r_o_function_of_B}. As can be seen from the figure the optimal ratio between the outer and inner radius deviate from the case of a cylinder of infinite length, especially for large values of $\langle B \rangle/B_\mathrm{rem}$.

\begin{figure}[!t]
  \centering
  \includegraphics[width=1\columnwidth]{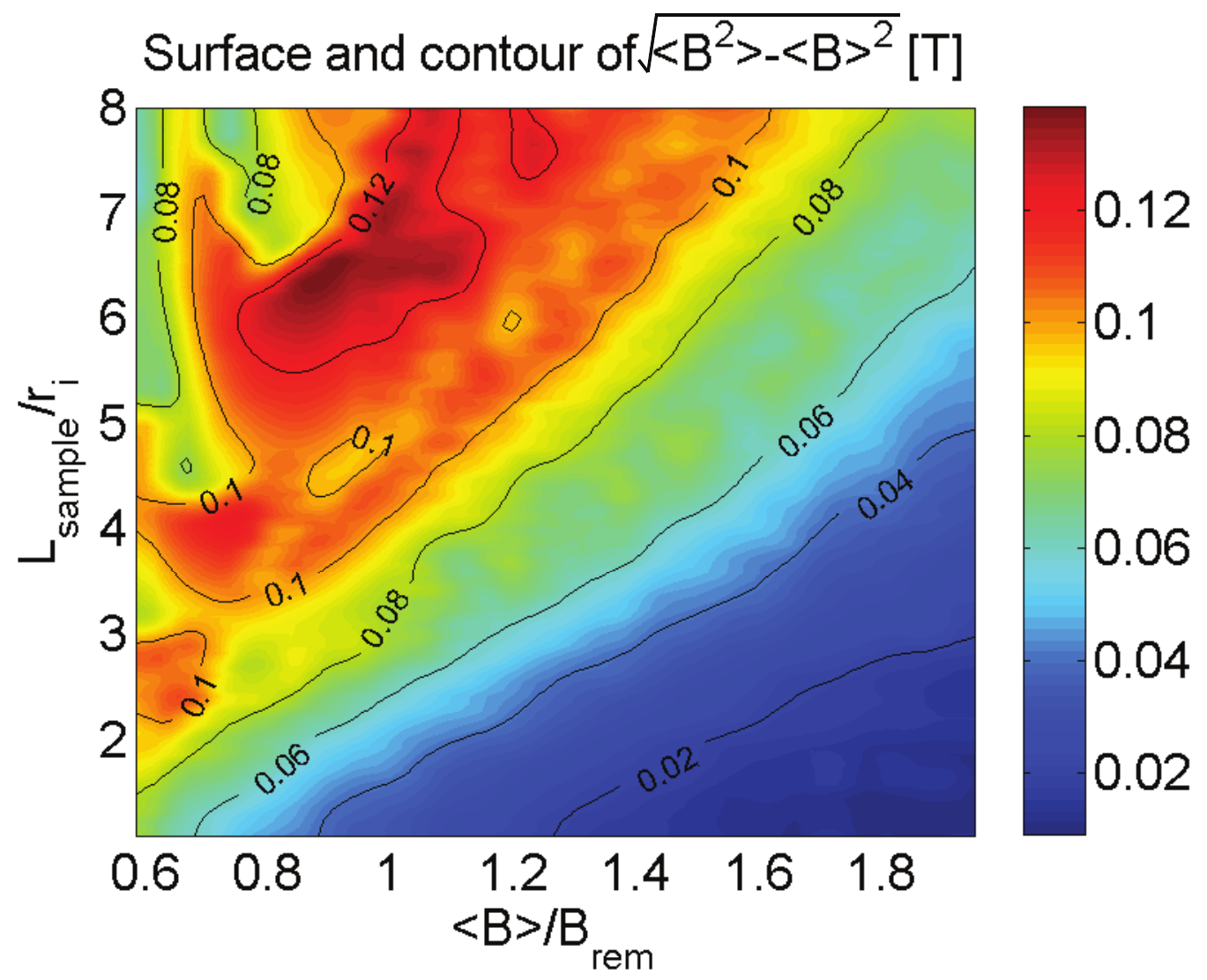}
      \caption{The standard deviation $\sqrt{\langle B^2 \rangle - \langle B \rangle ^2}$ for the optimal Halbach cylinders as a function of the ratio between the mean flux density and the remanence and the ratio between the length of the sample volume and the inner radius for the optimal configurations.}
      \label{Fig.Homogeneity_surface_image}
\end{figure}

It is also of importance to consider the homogeneity of the flux density in the cylinder bore. This can be characterized by the standard deviation, $\sqrt{\langle B^2 \rangle - \langle B \rangle ^2}$, which is shown for the optimal configurations in Fig. \ref{Fig.Homogeneity_surface_image} as a function of the flux density and the length of the sample volume, each normalized with respect to the remanence and the inner radius, respectively. By comparing with Fig. \ref{Fig.L_halbach_over_r_i_surface_image} it can be seen that the flux density in the sample volume is the most homogeneous, i.e. smallest, when $L_\n{Halbach}$ is much larger than $L_\n{sample}$. The inhomogeneous cases occur where $L_\n{Halbach}\simeq{}L_\n{sample}$ as is also expected as here flux density will ``leak'' out of the ends of the cylinder bore and thus the sample volume which will lower the homogeneity.

The optimal dimensions of a Halbach cylinder with a desired inner radius, length of the sample volume and mean flux density has thus been found and can be read directly of Figs. \ref{Fig.r_o_over_r_i_surface_image} and \ref{Fig.L_halbach_over_r_i_surface_image}. However, the Halbach cylinder magnet is not an equally efficient magnet design at all values of $B/B_\n{rem}$. This is discussed in the next section.

\section{The efficiency for a finite length Halbach cylinder}
A general figure of merit, $M^{*}$, used to characterize the efficiency of a magnet design is defined in Ref. \cite{Jensen_1996} as
\begin{eqnarray}\label{Eq.Mstar_definition}
M^{*}=\frac{\int_{V_\n{field}}||\mathbf{B}||^2dV}{\int_{V_\n{mag}}||\mathbf{B_\n{rem}}||^2dV}
\end{eqnarray}
where $V_\n{field}$ is the volume of the region where the magnetic field is created and $V_\n{mag}$ is the volume of the magnets. It can be shown that the maximum value of $M^{*}$ is 0.25, and a structure is considered reasonable efficient if it has $M^{*} \geq 0.1$ \cite{Jensen_1996}.

For a structure of infinite length with completely uniform remanence and magnetic flux density Eq. (\ref{Eq.Mstar_definition}) is reduced to
\begin{eqnarray}
M^{*}=\left(\frac{B}{B_\n{rem}}\right)^2\frac{V_\n{field}}{V_\n{mag}}~.
\end{eqnarray}

\begin{figure}[!t]
  \centering
   \includegraphics[width=1\columnwidth]{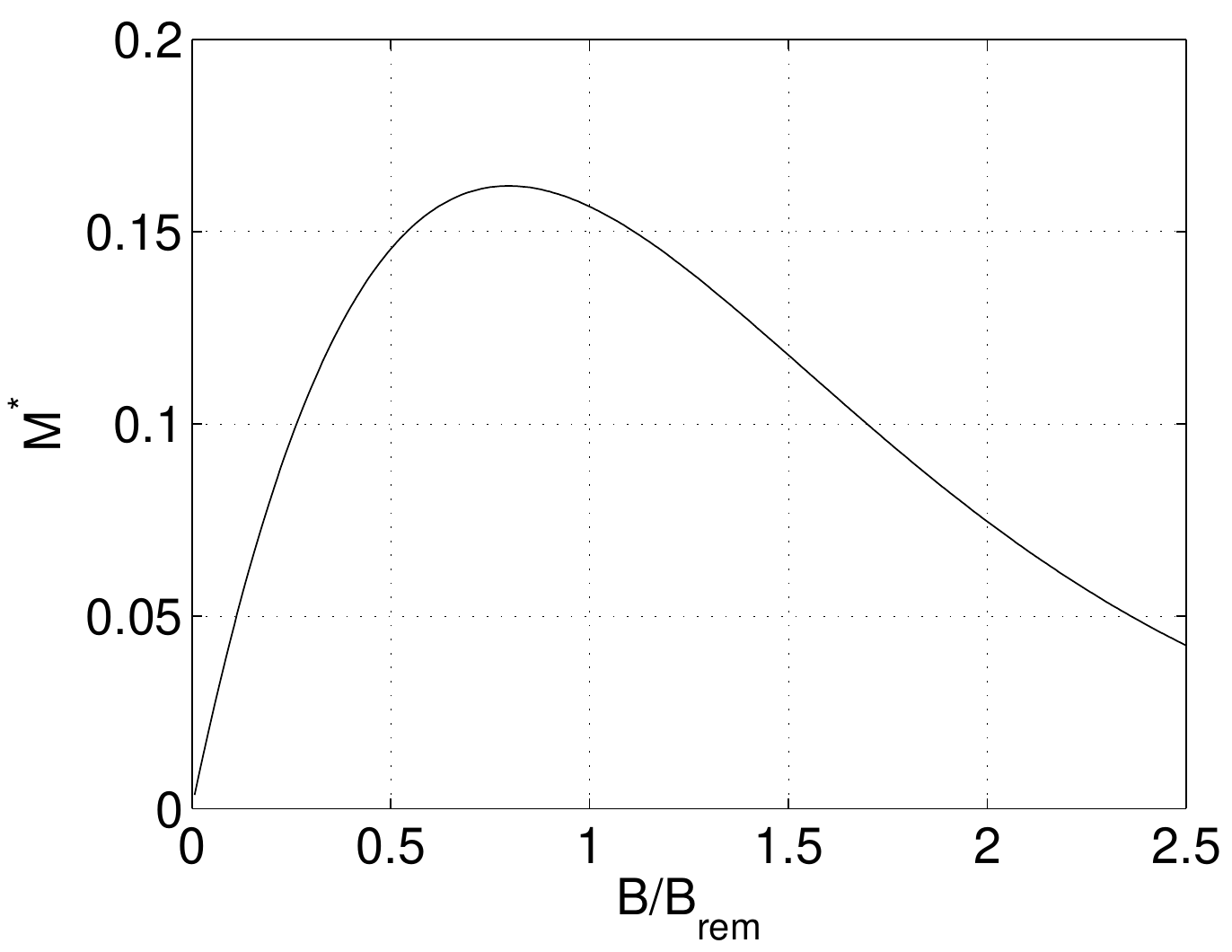}
      \caption{The $M^{*}$ parameter as a function of the ratio between the flux density and the remanence for a Halbach cylinder of infinite length.}
      \label{Fig.Mstar_for_Halbach}
\end{figure}

For a Halbach cylinder the figure of merit parameter can be found analytically for a cylinder of infinite length, through the relation for the flux density in the cylinder bore, Eq. (\ref{Eq.Halbach_flux_density}). Using this relation one gets \citep{Coey_2003}
\begin{eqnarray}\label{Eq.Mstar_Halbach}
M^{*} = \frac{\left(\frac{B}{B_\n{rem}}\right)^2}{e^{2B/B_\n{rem}}-1}~.
\end{eqnarray}
As shown in Ref. \cite{Abele_1990} for a Halbach cylinder of infinite length the maximum value of the figure of merit is $M^{*} \approx 0.162$ for a value of $B/B_\n{rem} \approx 0.797$, as can also be seen in Fig. \ref{Fig.Mstar_for_Halbach}. This means that the ideal ratio between the outer and the inner radius is given as $r_\n{o}/r_\n{i} = e^{B/B_\n{rem}} \approx 2.219$.

%The ideal ratio between the inner and outer radius can be found from the $M^*$ parameter. The $M^*$ parameter is given as
%\begin{eqnarray}
%M* &=& \left(\frac{||B||}{||B_\mathrm{rem}||}\right)^2\frac{V_\mathrm{bore}}{V_\mathrm{mag}}\\
%   &=& \left(\frac{||B_\mathrm{rem}||\mathrm{ln}\left(\frac{r_\mathrm{o}}{r_\mathrm{i}}\right)}{||B_\mathrm{rem}||}\right)^2\frac{\pi L r_\mathrm{i}^2}{\pi L (r_\mathrm{o}^2 - r_\mathrm{i}^2)}
%\end{eqnarray}
%where the expression for the flux density in the bore $||B|| = ||B_\mathrm{rem}||\mathrm{ln}\left(\frac{r_\mathrm{o}}{r_\mathrm{i}}\right)$ have been used.
%
%Defining the ratio between the outer and the inner radius as $x_r = \frac{r_\mathrm{o}}{r_\mathrm{i}}$ the above expression can be reduced to
%\begin{eqnarray}
%M* &=& \frac{\mathrm{ln}(x_r)^2}{x_r^2-1}
%\end{eqnarray}
%
%By differentiating this expression and setting it equal to zero the ideal ratio can be found. This results in the following equation which determines the optimal ratio
%\begin{eqnarray}
%\frac{\mathrm{ln}(x_r)x_r^2}{x_r^2-1} = 1
%\end{eqnarray}
%The solution to this equation are determined numerically as $x_r = 2.218$ for which $M^*$ obtains its maximum value of 0.162. This agree with the maximum value of $K$ reported by Ref. \cite{Coey_2003} who find $K \sim 0.80$ for $M* = 0.162$, as $x_r = e^{K}$.

\begin{figure}[!t]
  \centering
   \includegraphics[width=1\columnwidth]{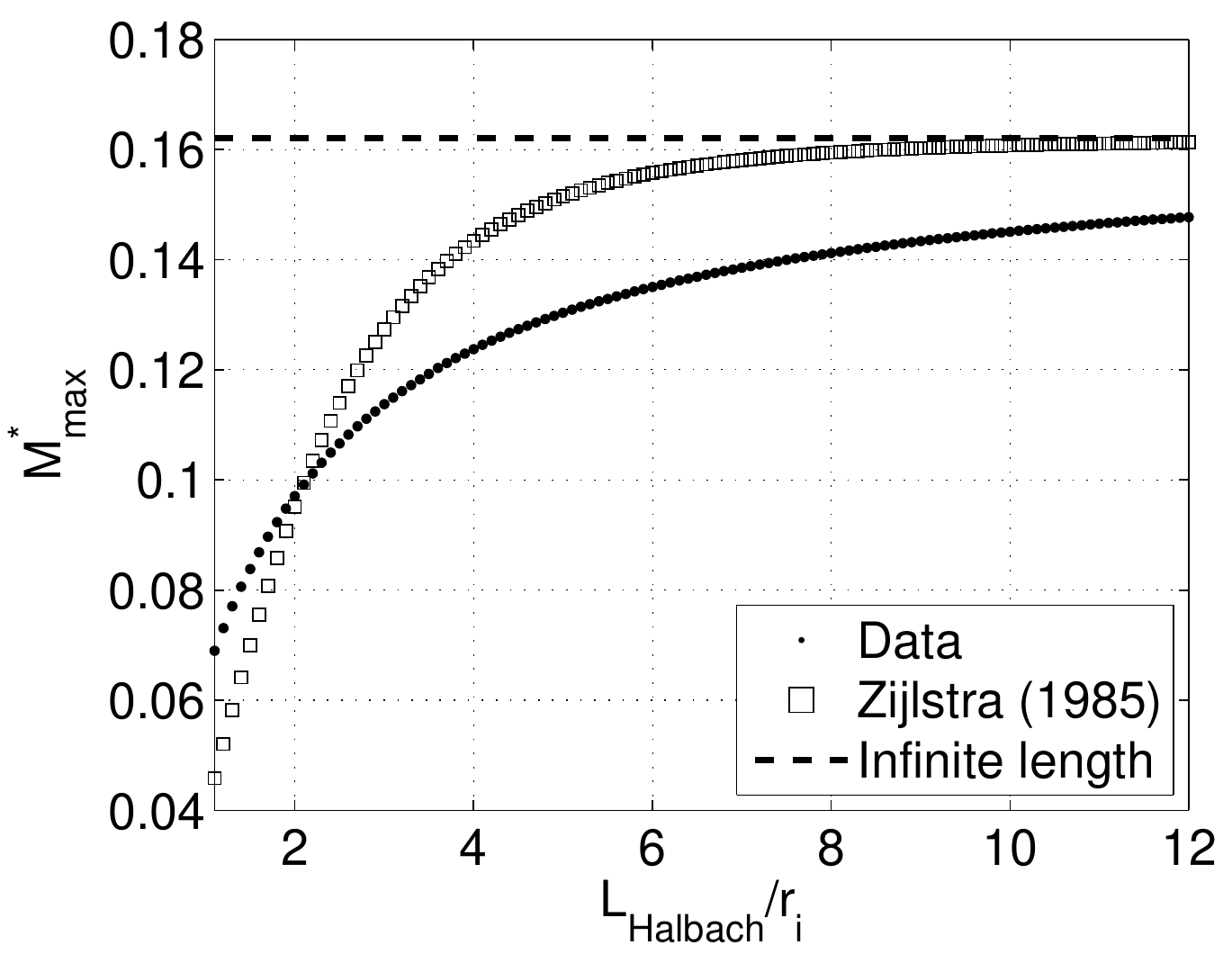}
      \caption{The maximum value of $M^*$ as a function of the ratio between $L_\n{Halbach}$ and the inner radius. The value in the case of a cylinder of infinite length is also shown.}
      \label{Fig.Mstarmax_function_of_L}
\end{figure}

However, for a cylinder of finite length the maximum value of $M^*$ will depend on the ratio between the length of the Halbach cylinder and the inner radius. The maximum value of $M^*$ has been found using the parameter variation simulations described above. Here, only the configurations where the length of the sample volume is equal to the length of the Halbach cylinder, i.e. $L_\n{Halbach} = L_\n{sample}$, are considered. The maximum value of $M^*$, found using spline interpolation, as a function of the ratio between $L_\n{Halbach}$ and the inner radius is shown in Fig. \ref{Fig.Mstarmax_function_of_L}. Also shown is the value for a cylinder of infinite length. It can be seen that even using a cylinder where the length is significantly greater than the inner radius the value of $M^*$ is far from the theoretical value.

Finally, the maximum value of $M^*$ for a finite length Halbach cylinder where only the field in the center of the Halbach cylinder is considered are shown. The field in the center of the Halbach has been calculated using the formula given in Ref. \cite{Zijlstra_1985}. Note that the value of $M^{*}$ calculated using the value of $B$ in the center of the Halbach cylinder assumes that this value of $B$ is present in all of the cylinder bore, which is not the case. Thus this calculation can significantly overestimate $<B>$ and hence $M^{*}$. The reason it has been included here is to shown that the proper mean value of the flux density in the cylinder bore must be calculated for a proper estimation of $M^{*}$.

\begin{figure}[!t]
  \centering
   \includegraphics[width=1\columnwidth]{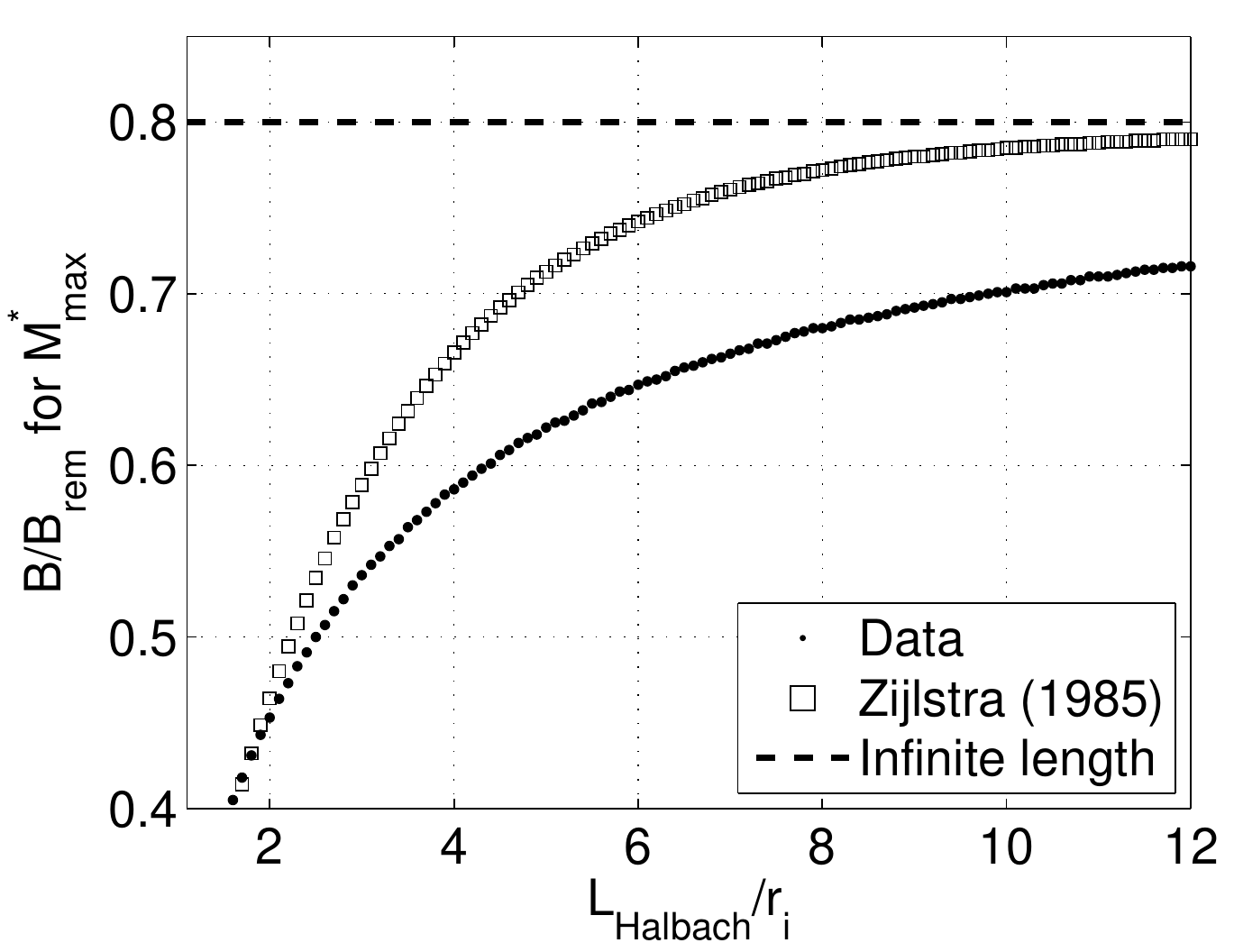}
      \caption{The value of $B/B_\n{rem}$ as a function of the ratio between $L_\n{Halbach}$ and the inner radius for the maximum values of $M^*$ shown in Fig. \ref{Fig.Mstarmax_function_of_L}. The value in the case of a cylinder of infinite length is also shown.}
      \label{Fig.Kmax_function_of_L}
\end{figure}

\begin{figure}[!t]
  \centering
   \includegraphics[width=1\columnwidth]{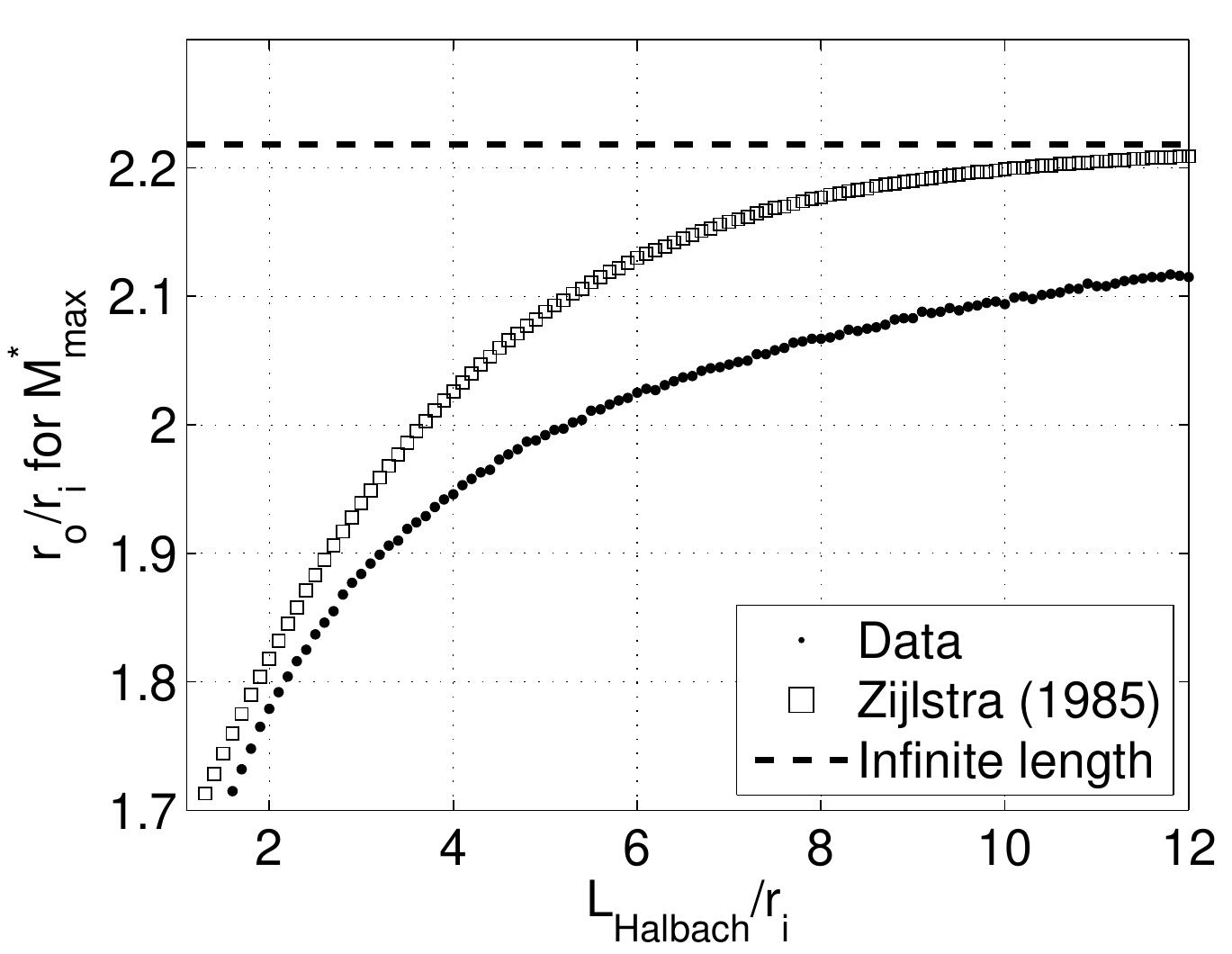}
      \caption{The ratio between the outer and inner radius as a function of the ratio between $L_\n{Halbach}$ and the inner radius for the maximum values of $M^*$ shown in Fig. \ref{Fig.Mstarmax_function_of_L}. The ratio in the case of a cylinder of infinite length is also shown.}
      \label{Fig.r_o_function_of_L_for_Mstarmax}
\end{figure}

The value of $B/B_\n{rem}$ and the value of the ratio between the outer and inner radius for the maximum values of $M^*$ are shown as functions of the ratio between $L$ and the inner radius in Figs. \ref{Fig.Kmax_function_of_L} and \ref{Fig.r_o_function_of_L_for_Mstarmax}. In both figures are also shown the theoretical value for a cylinder of infinite length as well as the expression for the flux density in the center of the sample volume, as given in Ref. \cite{Zijlstra_1985}. As was the case in Fig. \ref{Fig.Mstarmax_function_of_L} even a very long cylinder is far from the theoretical value.

\section{Conclusion}
The ideal, i.e. smallest, dimensions of a finite length Halbach cylinder that produces a given mean flux density in a specified sample volume has been found based on a numerical simulation using parameter variation. The dimensions have been reported in dimensionless units allowing for determination of the ideal dimensions for any given desired sample volume. Finally a general figure of merit efficiency parameter was calculated for the finite length Halbach cylinder as a function of the length of the cylinder. The dimensions of the most efficient Halbach cylinder were shown to only slowly approach the values for the case of a cylinder of infinite length as the length of the cylinder is increased.

\section*{Acknowledgements}
The author would like to acknowledge the support of the Programme Commission on Energy and Environment (EnMi) (Contract No. 2104-06-0032) which is part of the Danish Council for Strategic Research.


\newpage

\begin{thebibliography}{14}
\expandafter\ifx\csname natexlab\endcsname\relax\def\natexlab#1{#1}\fi

\bibitem{Mallinson_1973}
J. C. Mallinson, IEEE Trans. Magn. {\bf 9 (4)}, 678 (1973).

\bibitem{Halbach_1980}
K. Halbach, Nucl. Instrum. Methods{\bf 169} (1980).

\bibitem{Tura_2007}
A. Tura and A. Rowe, Proc. 2$^\mathrm{nd}$ Int. Conf. on Magn. Refrig. at Room Temp., 363 (2007).

\bibitem{Bjoerk_2010b}
R. Bj\o{}rk, C. R. H. Bahl, A. Smith, and N. Pryds, Int. J. Refrig. {\bf 33}, 437 (2010).

\bibitem{Moresi_2003}
G. Moresi and R. Magin, Concepts in Magn. Reson. Part B (Magn. Reson. Eng.), {\bf 19B}, 35 (2003).

\bibitem{Appelt_2006}
S. Appelt, H. K\"uhn, F. W H\"asing, and B. Bl\"umich, Nat. Phys. {\bf 2}, 105 (2006).

\bibitem{Sullivan_1998}
M. Sullivan, G. Bowden, S. Ecklund, D. Jensen, M. Nordby, A. Ringwall, and Z. Wolf, IEEE {\bf 3}, 3330 (1998).

\bibitem{Lim_2005}
J. K. Lim, P. Frigola, G. Travish, J. B. Rosenzweig, S. G. Anderson, W. J. Brown, J. S. Jacob, C. L. Robbins, and A. M. Tremaine, Phys. Rev. Spec. Top. - Accel. Beams, {\bf 8}, 072401 (2005).

\bibitem{Zhu_1993}
Z. Q. Zhu, D. Howe, E. Bolte, and B. Ackermann, IEEE Trans. Magn., {\bf 29}, 124--135 (1993)

\bibitem{Atallah_1997}
K. Atallah, D. Howe, and P. H. Mellor, Eighth Int. Conf. on Electr. Mach. Drives (Conf. Publ. No.444), 376 (1997).

\bibitem{Peng_2003}
Q. Peng, S. M. McMurry, and J. M. D. Coey, IEEE Trans. Magn., {\bf 39}, 1983 (2003).

\bibitem{Xia_2004}
Z. P. Xia, Z. Q. Zhu, and D. Howe, IEEE Trans. Magn., {\bf 40}, 1864 (2004).

\bibitem{Bjoerk_2010a}
R. Bj\o{}rk, A. Smith, and C. R. H. Bahl, J. Magn. Magn. Mater., 322:133--141,
  2010.

\bibitem{Mhiochain_1999}
T. R. {Ni Mhiochain}, D. Weaire, S. M. McMurry, and J. M. D. Coey, J. Appl. Phys., {\bf 86}, 6412 (1999).

\bibitem{Xu_2004}
X. N. Xu, D. W. Lu, G. Q. Yuan, Y. S. Han, and X. Jin, J. Appl. Phys., {\bf 95}, 6302 (2004).

\bibitem{Bjoerk_2008}
R. Bj\o{}rk, C. R. H. Bahl, A. Smith, and N. Pryds, J. Appl. Phys., {\bf 104}, 13910 (2008).

\bibitem{Bloch_1998}
F. Bloch, O. Cugat, G. Meunier, and J. C. Toussaint, IEEE Trans. Magn. {\bf 34}, 5 (1998).

\bibitem{Comsol_2005}
Comsol, Comsol Multiphysics Model Library, third ed. COMSOL AB, Chalmers Teknikpark 412 88 G (2005).

\bibitem{Schenk_2001}
O. Schenk, K. G\"{a}rtner, W. Fichtner, and A. Stricker, J. Future Gener. Comput. Syst. {\bf 18}, 69 (2001).

\bibitem{Schenk_2002}
O. Schenk, and K. G\"{a}rtner, Parallel Comput. {\bf 28} (2002).

\bibitem{Zimmermann_1993}
G. Zimmermann, J. Appl. Phys., {\bf 73}, 8436 (1993).

\bibitem{Zijlstra_1985}
H. Zijlstra, Phillips J. Res., {\bf 40}, 259 (1985).

\bibitem{Jensen_1996}
J. H. Jensen and M. G. Abele, J. Appl. Phys., {\bf 79}, 1157 (1996).

\bibitem{Coey_2003}
J. M. D. Coey and T. R. {Ni Mhiochain}, High Magnetic Fields (Permanent magnets), Chap. 2, p. 25, World Scientific (2003).

\bibitem{Abele_1990}
M. G. Abele and H. Rusinek, J. Appl. Phys., {\bf 67}, 4644 (1990).
\end{thebibliography}
\end{document}